\documentclass[a4,12pt]{article}
\usepackage{amsmath,amssymb,color,graphics,amscd,amsfonts,epsf,indentfirst}
\usepackage{epsfig}
\usepackage[dvipdfmx,hypertex]{hyperref}
\allowdisplaybreaks[4]

\makeatletter
\@addtoreset{equation}{section}

\makeatletter
\renewcommand\section{\@startsection {section}{1}{\z@}%
                                   {-3.5ex \@plus -1ex \@minus -.2ex}
                                   {2.3ex \@plus.2ex}%
                                   {\normalfont\large\bfseries}}
\renewcommand\subsection{\@startsection{subsection}{2}{\z@}%
                                     {-3.25ex\@plus -1ex \@minus -.2ex}%
                                     {1.5ex \@plus .2ex}%
                                    {\normalfont\bfseries}}

\def\btab{\begin{table}[h] \begin{center} \begin{tabular}{l lp{3in}}}
      \def\etab{\end{tabular} \end{center} \end{table}}
\def\btabm{\begin{center} \begin{tabular}}
    \def\etabm{\end{tabular} \end{center}}

\newcommand{\be}{\begin{equation}}
\newcommand{\ba}{\begin{eqnarray}}
\newcommand{\ea}{\end{eqnarray}}
\newcommand{\ee}{\end{equation}}

\def\del {\partial}

\begin{document}

\begin{titlepage}
  \thispagestyle{empty}

\begin{flushright}
  \end{flushright}
  \vspace{1cm}
  \begin{center}
    \font\titlerm=cmr10 scaled\magstep4
    \font\titlei=cmmi10 scaled\magstep4
    \font\titleis=cmmi7 scaled\magstep4
   
   
\centerline{\titlerm Observations on Open and Closed String \vspace{0.5cm}} \centerline{\titlerm Scattering Amplitudes at High Energies
\vspace{1.5cm}}
\noindent{{
       Pawe{\l} Caputa$\,^{a}\,$\footnote{e-mail:caputa@nbi.dk} and Shinji Hirano$\,^{b}\,$\footnote{e-mail:hirano@eken.phys.nagoya-u.ac.jp}
       }}\\
    \vspace{1.0cm}

    {\it $^{a}$The Niels Bohr Institute and The Niels Bohr International Academy,\\Blegdamsvej 17, DK-2100 Copenhagen \O, Denmark}
\vspace{0.5cm}

    {\it \it $^{b}$Department of Physics, Nagoya University, Nagoya 464-8602, Japan}

    \vspace{1cm}
    {\large \today}
  \end{center}

\vskip 5em

  \begin{abstract}
We study massless open and closed string scattering amplitudes in flat space at high energies. 
Similarly to the case of AdS space, we demonstrate that, under the T-duality map, the open string amplitudes are given by the exponential of minus minimal surface areas whose boundaries are cusped closed loops formed by lightlike momentum vectors.
We show further that the closed string amplitudes are obtained by gluing two copies of minimal surfaces along their cusped lightlike boundaries. 
This can be thought of as a manifestation of the Kawai-Lewellen-Tye (KLT) relation at high energies. 
We also discuss the KLT relation in AdS/CFT and its possible connection to amplitudes in ${\cal N}=8$ supergravity as well as the correlator/amplitude duality. 
  \end{abstract}

\end{titlepage}

 \noindent\rule\textwidth{.1pt}
 \vskip 2em \@plus 3ex \@minus 3ex

 \tableofcontents

 \vskip 2em \@plus 3ex \@minus 3ex
 \noindent\rule\textwidth{.1pt}
 \vskip 2em \@plus 3ex \@minus 3ex

\section{Introduction and Conclusions}

Minimal surfaces play important roles in string theory. 
For example, string scattering amplitudes are given by the path integrals weighted by the exponential of minus surface area. 
So the minimal surface gives the major contribution to the amplitudes. 
In fact, non-perturbative effects are similar in that they are given by the exponential of minus brane volumes. Thus the minimal surface of branes too yields the major contribution to the amplitudes.

\medskip
In this paper, among minimal surfaces, we are most concerned with those dominating massless string scattering amplitudes at high energies.
It is a well-known fact that high energy string scattering amplitudes are dominated by the saddle point of the string action \cite{Gross:1987kza}. 
The saddle point solutions represent minimal surfaces with spikes corresponding to vertex operator insertions.
Below we provide a simple but fresh look at this old problem. 

\medskip
Our study is by large motivated by Alday and Maldacena's work on gluon scattering in AdS/CFT  \cite{Alday:2007hr}.
According to their proposal, gluon scattering at strong coupling in ${\cal N}=4$ super Yang-Mills (SYM) is dual to massless open string scattering on D3-branes near the AdS horizon. Due to the large warping near the horizon, this is a high energy scattering process.
So the problem amounts to finding the minimal surfaces in $AdS_5$ space. However, it turns out to be difficult to directly find spiky minimal surfaces. The key to resolve this issue is to perform (formal) T-dualities under which the AdS space is self-dual. Since the T-dualities are expected to map massless momenta to lightlike windings, with the momentum conservation, this quandary gets mapped to Plateau's problem with cusped  closed  lightlike loops as the boundary. Not only did this trick provide technical ease, but also it led to new concepts such as the amplitude/Wilson loop duality \cite{Drummond:2007cf}.

\medskip
We take a lesson from string theory in AdS space to learn new perspectives on string theory in flat space. In fact, inspired by  \cite{Alday:2007hr}, Makeenko and Olesen advocated relevance of Douglas' approach to Plateau's problem \cite{Douglas} in the study of QCD scattering amplitudes in the Regge regime \cite{Makeenko:2008sh, Makeenko:2010dq}. We clarify that Douglas' method is nothing but the T-dual description of high energy string scattering in flat space. Moreover, we show that massless open string amplitudes at high energies are given by the exponential of minus minimal surface areas whose boundaries are cusped closed lightlike loops, similarly to the AdS case. We then make use of Douglas' method to study the KLT relation \cite{Kawai:1985xq} at high energies. This approach elucidates how two copies of minimal surfaces are glued along their cusped lightlike boundaries to yield massless closed string amplitudes.

\medskip
Finally, built on our observation on the KLT relation at high energies in the flat space, we discuss the KLT relation in AdS/CFT.
In particular, we suggest that amplitudes in ${\cal N}=8$ supergravity may be constructed by gluing two copies of minimal surfaces of Alday and Maldacena. We further argue that the correlator/amplitude duality proposed by \cite{Alday:2010zy} may be thought of as an incarnation of the KLT relation.

\section{String scattering amplitudes at high energies}

String scattering amplitudes are given by the sum of $\exp(-A)$ with the punctured areas $A$ specified by the scattering states. 
In the classical limit, they are dominated by the minimal surfaces $A_{\rm{min}}$. The quantum numbers of amplitudes are momenta $k_i^{\mu}$ of the scattering states. When they are large, the classical approximation becomes accurate. In fact the exponential factor of the scattering amplitudes has the effective action
\begin{equation}
S={1\over 4\pi\alpha'}\int d\tau d\sigma \del_a X^{\mu}\del^a X_{\mu}+i\sum_{i=1}^nk_i^{\mu}X_{\mu}(\sigma_i)\ ,
\end{equation}
where the second term is the center-of-mass part of vertex operators. 
By scaling $X^{\mu}$ to $\alpha'|k|X^{\mu}$, it is easy to see that the high energy scattering can be approximated by the saddle point of the action $S$. Intuitively, at high energies strings are stretched so long that their oscillations become negligible. Hence high energy string scattering amplitudes are insensitive to the types of asymptotic states. 
For our purpose, however, we will restrict ourselves to massless string states.

\medskip
In the closed string case, the saddle point is the solution to the Laplace equation
\begin{equation}
\del_a\del^aX^{\mu}=2\pi i\alpha'\sum_{i=1}^nk_i^{\mu}\delta^2(\sigma-\sigma_i)\ .
\end{equation}
This can be easily solved to
\begin{align}
X^{\mu}={i\alpha'\over 2}\sum_{i=1}^nk_i^{\mu}\log|z-z_i|^2\ ,
\end{align}
where $z=\sigma+i\tau$. 
The saddle point action then yields
\begin{equation}
S_{\rm{min}}=-{\alpha'\over 2}\sum_{i,j=1}^nk_i\cdot k_j\log|z_i-z_j|\ .
\end{equation}
Thus the high energy scattering amplitude, up to the polarization dependence, takes the form
\begin{equation}
A^{\rm{cl}}_n\sim\int \prod_{i=2}^{n-2}d^2z_i\exp\left(-S_{\rm{min}}\right)=\int \prod_{i=2}^{n-2}d^2z_i\prod_{\stackrel{i,j=1}{i<j}}^n|z_i-z_j|^{\alpha' k_i\cdot k_j}\ .\label{clAn}
\end{equation}
By using the conformal Killing group, one can fix three of the vertex operator positions to $(z_1, z_{n-1}, z_n)=(0, 1, \infty)$. 
There is no need for carrying out the moduli integrations over $z_i$'s. They are well approximated by the saddle point value. 
This amplitude of course coincides with the Virasoro-Shapiro amplitude at $\alpha' |k_i\cdot k_j| \gg 1$ where the intercept can be neglected.

\medskip
Similarly, in the open string case, one finds\footnote{The factor of two in the exponent is due to image charges and the restriction to the upper half-plane (UHP).} 
\begin{equation}
A^{\rm{op}}_n\sim\int \prod_{i=2}^{n-2}d\sigma_i\prod_{\stackrel{i,j=1}{i<j}}^n|\sigma_i-\sigma_j|^{2\alpha' k_i\cdot k_j} + \rm{cyclic},
\label{opAn}
\end{equation}
where \lq\lq cyclic" stands for the sum over cyclic orderings of vertex operators on the disc.
We fix three of the vertex operator positions to $(\sigma_1, \sigma_{n-1}, \sigma_n)=(0, 1, \infty)$. The moduli integrals over $\sigma_i$'s are again dominated by the saddle point. 
This amplitude coincides with the Veneziano amplitude and its Koba-Nielsen generalization at $\alpha' |k_i\cdot k_j| \gg 1$ where the intercept can be neglected.

\medskip
In the 4-point case, the closed string amplitude (\ref{clAn}) becomes
\begin{equation}
A^{\rm{cl}}_4\sim\int d^2z_2|z_2|^{-\alpha' s/2}|1-z_2|^{-\alpha' t/2}\ ,
\end{equation}
where the Mandelstam variables $s$, $t$, and $u=-(s+t)$ are defined in Appendix \ref{Ap:Kinematics}. The saddle point of the integral is given by
\begin{equation}
z_2=-{s\over u}\ .
\end{equation}
This yields the well-known soft exponential behavior of the hard scattering
\begin{equation}
A^{\rm{cl}}_4\sim\exp\left[-{\alpha'\over 2}\left(s\ln s+t\ln t+u\ln u\right)\right]\ .
\label{cl4ptGM}
\end{equation}
Similarly, the 4-point open string amplitude is given by
\begin{equation}
A^{\rm{op}}_4\sim\int d\sigma_2|\sigma_2|^{-\alpha' s}|1-\sigma_2|^{-\alpha' t}
\sim \exp\biggl[-\alpha'\left(s\ln s+t\ln t+u\ln u\right)\biggr]\ .
\label{op4ptGM}
\end{equation}

\medskip
We have so far reviewed the standard understanding of string scattering amplitudes at high energies by Gross and Mende \cite{Gross:1987kza}. (This soft behavior for fixed-angle scattering was already observed by Veneziano in his celebrated paper \cite{Veneziano:1968yb}.) 
In the next section, motivated by the work of Alday and Maldacena on gluon scattering amplitudes in AdS/CFT \cite{Alday:2007hr} , we will provide a fresh perspective on high energy string scattering in flat space.

\section{T-duality and cusped lightlike loops}

In \cite{Alday:2007hr}
the massless {\it open} string scattering in AdS space was considered as a dual of gluon scattering amplitudes in ${\cal N}=4$ SYM at strong coupling. Strictly speaking, the S-matrix in CFT is not well-defined, since there is no notion of asymptotic states due to the scale invariance. However, by introducing the IR cutoff, it is possible to formally create asymptotic states to scatter.
On the AdS side, this corresponds to putting D3-branes near the horizon and scattering open string states on them. 
Since the proper momenta of open strings near the AdS horizon are very large due to the warp factor, this is a high energy scattering process.
Accordingly, the amplitudes are well approximated by the saddle points.

\medskip
However, it is very hard, if not impossible, to directly find the saddle points in AdS space with massless open string vertex insertions. 
Fortunately, there is a clever way to circumvent this difficulty. 
Namely, by performing (formal) T-dualities, massless momenta become lightlike windings, and the AdS horizon is mapped to the AdS boundary. Consequently, the vertex insertions become a closed lightlike contour on the AdS boundary. 
This way the problem is mapped to that of finding minimal surfaces with cusped closed loops formed by massless momentum vectors.\footnote{This also implies a duality between MHV amplitudes and lightlike Wilson loops. The T-duality is a strong-weak coupling duality with respect to $\alpha'\sim1/\sqrt{\lambda}$ where $\lambda$ being 't Hooft coupling. However, this duality turns out to hold order by order in perturbation theory of ${\cal N}=4$ SYM \cite{Drummond:2007cf}.}

\begin{figure}[h!]
\centering
\includegraphics[height=2.5in]{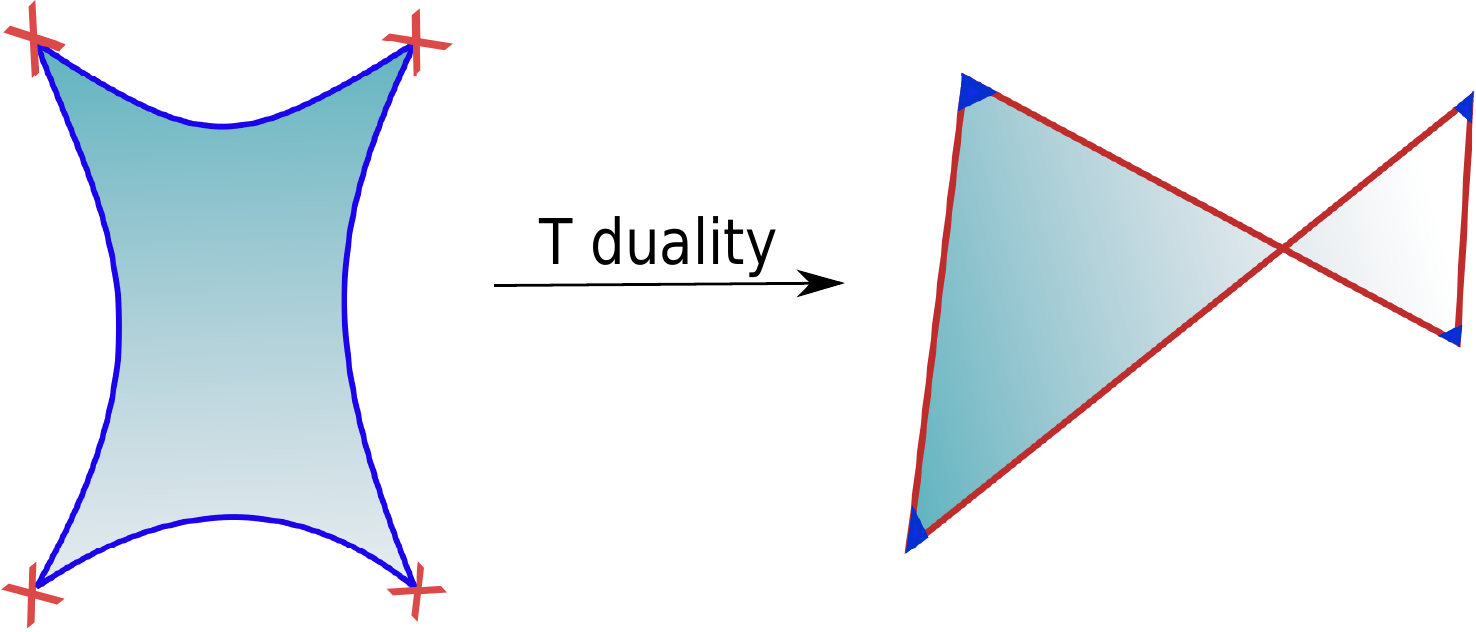}
\caption{The vertex insertion points are mapped to the lightlike segments, and the intervals between them are mapped to the cusps.}
\label{fig.tdual}
\end{figure}

\medskip
This suggests that a similar T-dual description may exist for massless open string scattering in flat space at high energies. 
We now demonstrate that this is indeed the case: As reviewed above, the saddle point solution is given by
\begin{align}
X^{\mu}=2i\alpha'\sum_{i=1}^nk_i^{\mu}\log\sqrt{(\sigma-\sigma_i)^2+\tau^2}\ .
\end{align}
The T-dual coordinates $Y^{\mu}$ are related to the original coordinates $X^{\mu}$ by \cite{Buscher:1987qj}
\begin{align}
\del_{\tau}Y^{\mu}&=\del_{\sigma}X^{\mu}\ ,\nonumber\\
\del_{\sigma}Y^{\mu}&=-\del_{\tau}X^{\mu}\ .
\end{align}
Note that these are the Cauchy-Riemann equations between the complex coordinates $Z^{\mu}=Y^{\mu}+iX^{\mu}$ and $z=\sigma+i\tau$.
Thus the T-dual coordinates can be found as
\begin{equation}
Y^{\mu}=-4i\alpha'\mbox{Im}\left[\sum_{i=1}^nk_i^{\mu}\log\sqrt{(\sigma-\sigma_i)+i\tau}\right]
=-2i\alpha'\sum_{i=1}^nk_i^{\mu}\arctan\left({\sigma-\sigma_i\over \tau}\right)\ .
\end{equation}
Approaching to the boundary $\tau=0$, the T-dual coordinates become the step function
\begin{equation}
Y^{\mu}\stackrel{\tau\to 0}{\longrightarrow} -2\pi i\alpha'\sum_{i=1}^nk_i^{\mu}\theta(\sigma-\sigma_i)\ .\label{LLprofile}
\end{equation}
As promised, {\it this describes the closed lightlike loop composed of massless momentum vectors $\vec{k}_i$'s with the momentum conservation $\sum_{i=1}^n\vec{k}_i=0$.}\footnote{ Similar observation was previously made in \cite{Fairlie:2008dg}}
This shows manifestly that the vertex insertion points are mapped to the lightlike segments, and the intervals between them are mapped to the cusps, as depicted in Figure \ref{fig.tdual}.\footnote{We thank Masaki Shigemori for discussions on this point.}

\medskip
The minimal area with this boundary condition can be evaluated by the T-dual action:
\begin{align}
S&\equiv{1\over 4\pi\alpha'}\int_{-\infty}^{\infty} d\sigma\int_0^{\infty}d\tau \del_aY^{\mu}\del^aY_{\mu}=-{1\over 4\pi\alpha'}\int_{-\infty}^{\infty}
d\sigma\lim_{\tau\to 0}Y^{\mu}\del_{\tau}Y_{\mu}
\nonumber\\
&=-\alpha'\sum_{i,j=1}^nk_i\cdot k_j\log(\sigma_i-\sigma_j)\ .\label{LLaction}
\end{align}
Hence the T-dual amplitude indeed reproduces (\ref{opAn}).

\medskip
We reiterate the main message that at high energies massless open string amplitudes in flat space are given by the exponential of minus minimal surfaces whose boundaries are cusped closed loops formed by lightlike momentum vectors, similarly to the AdS case.

\subsection{Douglas' method}

Finding minimal surfaces with given boundary conditions is known as Plateau's problem in mathematics and solved by Jesse Douglas in 1931 \cite{Douglas}. In his approach, he introduced a functional, now called the Douglas integral or Douglas functional, and found minimal surfaces by extremizing it.
In physics literature, the Douglas functional appeared, for example, in \cite{Migdal:1993sx}. More recently, it was extensively used  in a series of papers by Makeenko and Olesen in the study of QCD scattering amplitudes \cite{Makeenko:2008sh,Makeenko:2010dq}. 

\medskip
The T-dual computation of high energy scattering amplitudes is equivalent to Douglas' method. 
In fact, as we will see now, it amounts to the minimization problem of the Douglas functional 
\begin{equation}
D\left[C\right]:=\intop_{-\infty}^{+\infty}d\zeta_{1}\intop_{-\infty}^{+\infty}d\zeta_{2}\,\del_{\zeta_1}C(\zeta_{1})
\cdot \del_{\zeta_2}C(\zeta_{2})\:\log
\left(\sigma(\zeta_{1})-\sigma(\zeta_{2})\right)\ ,
\end{equation}
where $C$ denotes the boundary profile, and $\zeta(\sigma)$ is the reparametrization on the boundary. To elaborate on it further,
first note that in flat space the target space coordinates with the boundary profile $C$ are given by
\begin{align}
Y^{\mu}(\sigma,\tau)={1\over\pi}\intop_{-\infty}^{+\infty}d\sigma'{C^{\mu}\left(\zeta(\sigma')\right)\tau\over (\sigma-\sigma')^2+\tau^2}\ .
\label{Dirichlet}
\end{align}
The T-dual action evaluated on this solution then yields
\begin{align}
S={D[C]\over 4\pi^2\alpha'}\ .\label{Dougaction}
\end{align}
Note that since the Dirichlet solution (\ref{Dirichlet}) does not preserve the boundary reparametrization, this degrees of freedom must be integrated in the path integral in order to ensure the diffeomorphism invariance. Put differently, the Virasoro constraints are respected only for the saddle point values of $\sigma(\zeta_i)$'s. The minimal area is given by the action (\ref{Dougaction}) with the saddle point values of $\sigma(\zeta_i)$'s: 
\begin{align}
\sum_{j=1}^n{C'(\zeta_i)\cdot C'(\zeta_j)\over \sigma(\zeta_i)-\sigma(\zeta_j)}=0\ .
\label{Minimization}
\end{align}
When the boundary profile $C^{\mu}$ is given by (\ref{LLprofile}), the Douglas functional (\ref{Dougaction}) yields (\ref{LLaction}). It then follows that the saddle point value of the Douglas functional gives open string high energy scattering amplitudes.

\medskip
Hence we have shown that Douglas' method advocated in \cite{Makeenko:2008sh,Makeenko:2010dq} is nothing but the T-dual description of high energy open string scattering amplitudes.

\subsection{Lower point amplitudes}

As examples, we give more explicit forms of the Douglas functional corresponding to the 4, 5, and 6-point amplitudes. In the 4-point case, we have the Douglas functional
\begin{align}
D_{4}=\left(2\pi\alpha'\right)^2\left[-s\,\log\left(\lambda_{1}\right)-t\,\log\left(1-\lambda_{1}\right)\right]\ ,
\label{eq:D4lambda}
\end{align}
where $\lambda_1$ is the cross ratio $\frac{\sigma_{12}\sigma_{34}}{\sigma_{13}\sigma_{24}}$ with the notation $\sigma_{ij}=\sigma_i-\sigma_j$. Hence the amplitude is given by
\begin{align}
A_4=\int_0^1 d\lambda_1\lambda_1^{-\alpha' s}(1-\lambda_1)^{-\alpha' t}
\sim \exp\biggl[-\alpha'\left(s\ln s+t\ln t+u\ln u\right)\biggr]\ .
\end{align}
with the saddle point $\lambda_1=-s/u$. Note that the saddle point equations \eqref{Minimization} simplify considerably when expressed in terms of cross-ratios as shown in Appendix \ref{Ap:Saddle}.

\medskip
In the 5-point case, we find
\begin{align}
D_{5}=\left(2\pi\alpha'\right)^2&\left[-s_{1}\log\left(\lambda_1\right)-t_{1}\log\left(1-\lambda_1\right)-s_{2}\log\left(\lambda_2\right)-t_{2}\log\left(1-\lambda_2\right)\right.\\
&\left.-(t-t_1-t_2)\log\left(1-\lambda_1 \lambda_2\right)\right]\ ,\nonumber
\label{D5l}
\end{align}
where $\lambda_1$ and $\lambda_2$ are two cross ratios $\frac{\sigma_{12}\sigma_{35}}{\sigma_{13}\sigma_{25}}$ and $\frac{\sigma_{13}\sigma_{45}}{\sigma_{14}\sigma_{35}}$ respectively. The generalized Mandelstam variables are defined  in Appendix \ref{Ap:Kinematics}.
This yields the amplitude
\begin{align}
A_5=\int_0^1d\lambda_1\int_0^1d\lambda_2
\lambda_{1}^{-\alpha' s_{1}}(1-\lambda_{1})^{-\alpha' t_{1}}\lambda_{2}^{-\alpha' s_{2}}(1-\lambda_{2})^{-\alpha' t_{2}}(1-\lambda_{1}\lambda_{2})^{-\alpha' t+\alpha' t_{1}+\alpha' t_{2}}\ .
\end{align}
This is a generalization of the Veneziano amplitude that was first written down by Bardakci and Ruegg \cite{Bardakci:1969mg}.
As a side remark, there is a special kinematic regime $t\sim t_1+t_2$ in which the 5-point amplitude factorizes simply to two 4-point amplitudes. 
The saddle point can be found explicitly. But it is rather involved and not particularly illuminating. So we will not show it here.

\medskip
In the 6-point case, we find with a little effort 
\begin{align}
D_6=&\left(2\pi\alpha'\right)^2\left[-s_1\log\left(\lambda_1\right)-s_2\log\left(\lambda_2\right)-s_3\log\left(\lambda_3\right)+t_1\log\left(\frac{1-\lambda_1\lambda_2}{1-\lambda_1}\right)\right.\nonumber\\
&+t_2\log\frac{(1-\lambda_1\lambda_2)(1-\lambda_2\lambda_3)}{(1-\lambda_2)(1-\lambda_1\lambda_2\lambda_3)}+t_3\log\left(\frac{1-\lambda_2\lambda_3}{1-\lambda_3}\right)-t\log\left(1-\lambda_1\lambda_2\lambda_3\right)\\
&\left.+\Sigma_1\log\left(\frac{1-\lambda_1\lambda_2\lambda_3}{1-\lambda_1\lambda_2}\right)+\Sigma_2\log\left(\frac{1-\lambda_1\lambda_2\lambda_3}{1-\lambda_2\lambda_3}\right)\right]\ ,\nonumber
\end{align}
where the cross ratios are defined by $\lambda_{1}=\frac{\sigma_{12}\sigma_{36}}{\sigma_{13}\sigma_{26}}$, $\lambda_{2}=\frac{\sigma_{13}\sigma_{46}}{\sigma_{14}\sigma_{36}}$, and $\lambda_{3}=\frac{\sigma_{14}\sigma_{56}}{\sigma_{15}\sigma_{46}}$.
The amplitude takes the form
\begin{align}
A_6=\int_0^1\!d\lambda_1&\!\int_0^1\!d\lambda_2 \!\int_0^1\!\!d\lambda_3
\lambda^{-\alpha' s_1}_{1}(1-\lambda_1)^{-\alpha' t_1}\lambda^{-\alpha' s_2}_{2}(1-\lambda_2)^{-\alpha' t_2}\lambda^{-\alpha' s_3}_{3}(1-\lambda_3)^{-\alpha'  t_3}\\
&\times (1-\lambda_{1}\lambda_{2})^{-\alpha' \Sigma_{1}+\alpha' t_{1}+\alpha' t_{2}}(1-\lambda_2 \lambda_3)^{-\alpha' \Sigma_{2}+\alpha' t_{2}+\alpha' t_{3}}(1-\lambda_{1}\lambda_{2}\lambda_{3})^{-\alpha' t-\alpha' t_{2}+\alpha' \Sigma_1+\alpha' \Sigma_{2}}\ .\nonumber
\end{align}
Again this is the Bardakci and Ruegg formula for six scalars \cite{Bardakci:1969mg}.
 Similarly to the 5-point case, as a side remark, simple factorizations occur in special kinematic regimes.
The saddle point can be found explicitly but is not so illuminating to be presented.
It is straightforward but becomes increasingly tedious to find the explicit form of the Douglas functionals for higher point amplitudes.

\section{Closed string amplitudes and KLT relation}

Closed string amplitudes are related to open string amplitudes by \lq\lq squaring" the latter, since vertex operators of the former $V^{\rm{cl}}(z,\bar{z})$ are formally constructed from the latter $V^{\rm{op}}(z)$ by $V^{\rm{cl}}(z,\bar{z})=V^{\rm{op}}(z)V^{\rm{op}}(\bar{z})$. This is known as the KLT relation  \cite{Kawai:1985xq} (see also a recent review \cite{Sondergaard:2011iv}).
In particular, $n$-point graviton amplitudes can be constructed from $n$-point gluon amplitudes via the KLT relation.
In the 4-point case, the Virasoro-Shapiro amplitude is given by a product of two Veneziano amplitudes \cite{Polchinski}
\begin{equation}
A^{\rm{cl}}_{4}\left(\alpha',s,t\right)=\kappa^{2}\alpha' \sin\left(\frac{\pi}{2}\alpha' k_{2}\cdot k_{3}\right)A^{\rm{op}}_{4}\left(\frac{\alpha'}{4}s,\frac{\alpha'}{4}t\right) \overline{A_{4}^{\rm{op}}}\left(\frac{\alpha'}{4}s,\frac{\alpha'}{4}t\right),
\end{equation}
where $\kappa$ is some constant. 
Note that the KLT relation involves the rescaling of $\alpha'$ by the factor of $1/4$ in open string amplitudes.

\begin{figure}[h!]
\centering
\includegraphics[height=1.8in]{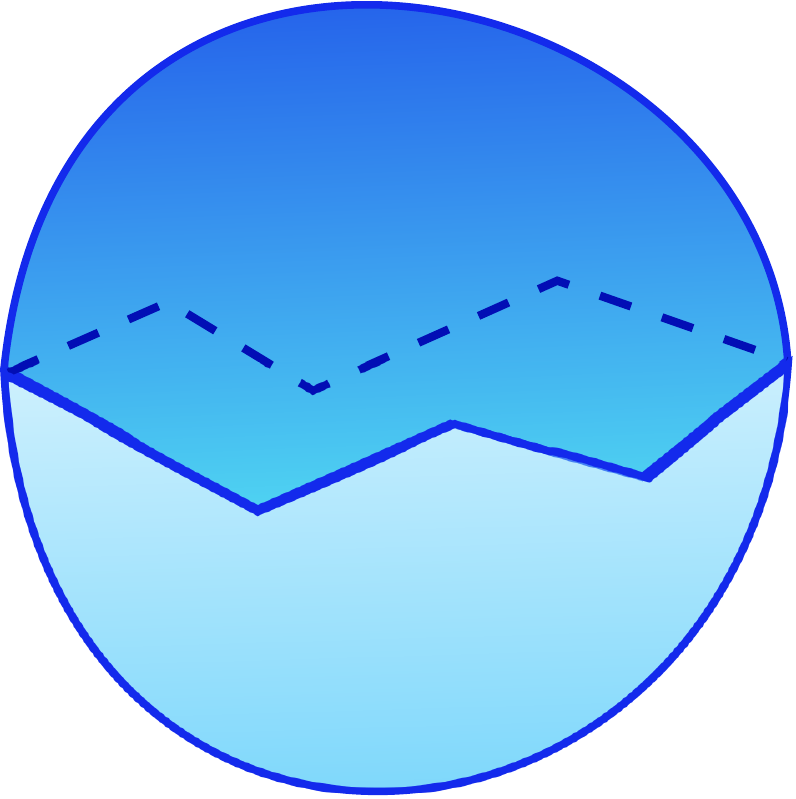}
\caption{Two disks glued along the cusped closed lightlike boundary to form a sphere. In flat space the actual sphere is flattened. }
\label{fig.KLT}
\end{figure}

\medskip
We are now interested in the KLT relation at high energies and particularly the interpretation of it. 
Since at high energies amplitudes are given by the exponential of minus surface areas, one may expect that the KLT relation becomes simply
\begin{equation}
A^{\rm{cl}}(\alpha')\sim \exp\left({-A_{S^2}}\right)= \exp\left({-(A_{D_2}+\overline{A_{D_2}})}\right)
\sim A^{\rm{op}}(\alpha'/4)\overline{A^{\rm{op}}}(\alpha'/4)\ .
\end{equation}
In the massless case, this has the interpretation that high energy graviton amplitudes are obtained by gluing two copies of minimal surfaces along the cusped lightlike boundary to form a sphere, as shown in Figure \ref{fig.KLT}.\footnote{It is straightforward to generalize this to the massive case.}
The Douglas functional approach elucidates how two copies of minimal surfaces are glued along their cusped lightlike boundaries to yield massless closed string amplitudes.
Using the Douglas functional, we can glue two surfaces of disk topology as
\begin{align}
-4\pi^2\alpha' A_{S^2}=\intop^{+\infty}_{-\infty}d\zeta_{1}\intop^{+\infty}_{-\infty}d\zeta_{2}\,
C'(\zeta_1)\cdot C'(\zeta_2)\biggl(\log\left(\sigma(\zeta_1)-\sigma(\zeta_2)\right)
+\log\left(\overline{\sigma}(\zeta_1)-\overline{\sigma}(\zeta_2)\right)\biggr)\ ,
\end{align}
where $C^{\mu}(\zeta)$ is given by \eqref{LLprofile} and $\alpha'$ to be replaced by $\alpha'/4$. 
This indeed yields the graviton amplitude at high energies
\begin{align}
A^{\rm{cl}}_n\sim\exp\left(-A_{S^2}\right)=\int \prod_{i=2}^{n-2}d^2z_i\prod_{\stackrel{i,j=1}{i<j}}^n|z_i-z_j|^{\alpha' k_i\cdot k_j}\ .
\label{KLTdoug}
\end{align}
Note that since the saddle point solutions of $z_i$'s and $\bar{z}_i$'s are the same, we have 
\begin{align}
A_{S^2}=A_{D_2}+\overline{A_{D_2}}=2A_{D_2}\ .
\end{align}
As an example, one can easily see that the 4-point massless amplitudes at high energies (\ref{cl4ptGM}) and (\ref{op4ptGM}) indeed obey the relation
\begin{align}
A^{\rm{cl}}_4(\alpha')\sim \exp\left({-A_{S^2}}\right)= \exp\left({-2A_{D_2}}\right)
\sim A^{\rm{op}}_4(\alpha'/4)^2\ .\label{4ptsquare}
\end{align}
It is clear that this KLT relation holds for generic $n$-point amplitudes at high energies.

\section{KLT relation in AdS/CFT?}

We have learned in the previous sections that 
(1) high energy massless open string amplitudes are given by the exponential of minus minimal surface areas with lightlike polygonal boundary, and 
(2) high energy graviton amplitudes can be constructed by gluing two copies of these minimal surfaces along the lightlike boundary.
\subsection{$\mathcal{N}=8$ amplitudes from AdS/CFT?}

\begin{figure}[h!]
\centering
\includegraphics[height=2.6in]{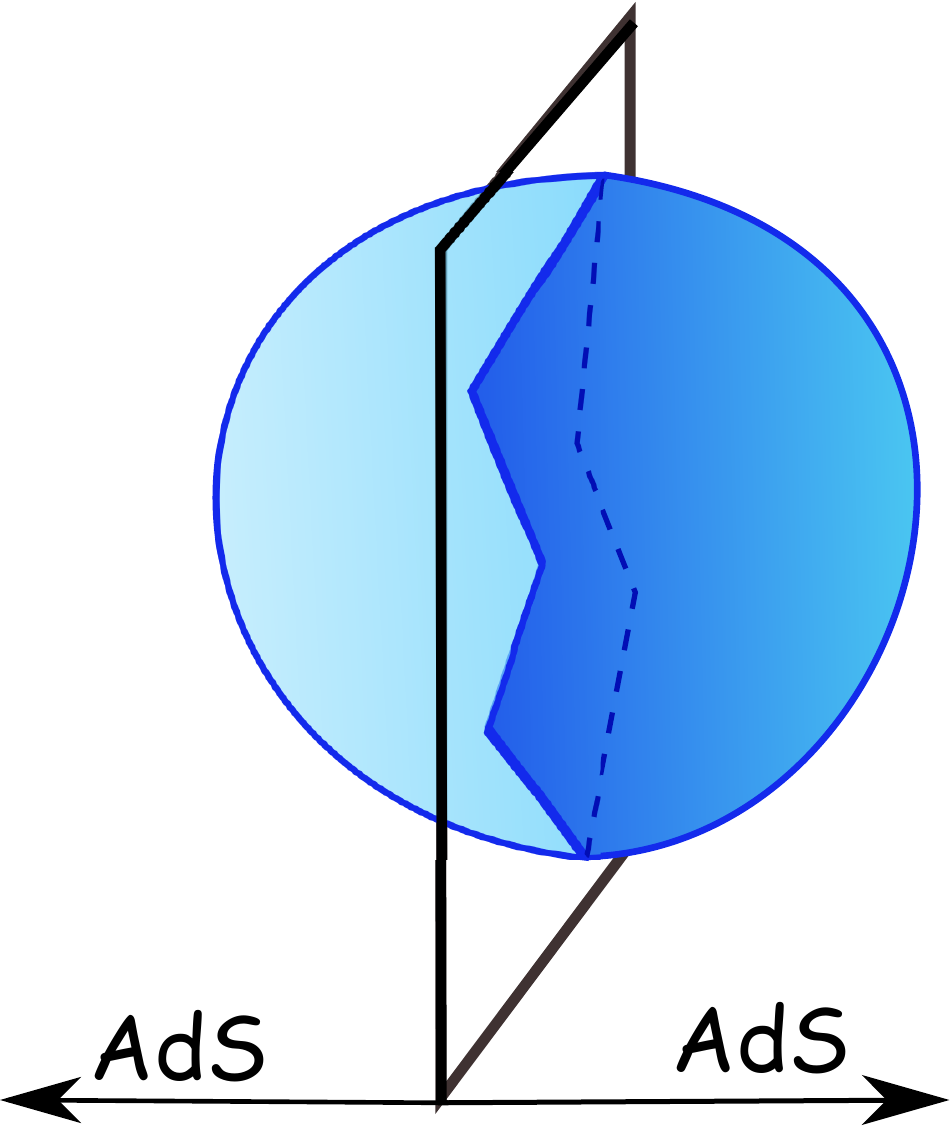}
\caption{Two copies of Alday-Maldacena's minimal surfaces glued to yield graviton scattering amplitudes at high energies in AdS via the KLT relation.}
\label{fig.glue}
\end{figure}

As mentioned earlier, in string theory on $AdS_5\times S^5$, Alday and Maldacena found the T-dual minimal surfaces for gluon scattering in $\mathbb{R}^{1,3}\subset AdS_5$ \cite{Alday:2007hr}. They are dual to gluon scattering in ${\cal N}=4$ SYM at strong coupling. 
We stress that the gluon scattering in ${\cal N}=4$ SYM is a high energy process in dual string theory on $AdS_5\times S^5$.
Thus we expect, from the string theory viewpoint, that there is a simple KLT relation similar to the flat space case. 
Namely, we may glue two copies of these minimal surfaces along their boundaries to obtain the amplitude of graviton scattering along the cutoff D3-branes in $AdS_5$. Note that, as shown in Figure \ref{fig.glue}, we double the AdS space too in gluing two minimal surfaces.  
Then the question is: What is this dual to?

\medskip
Gluing two copies of minimal surfaces corresponds to squaring lightlike Wilson loops of two copies of ${\cal N}=4$ SYM's.
This is equivalent to squaring (all-loop, tree-amputated) MHV amplitudes of two copies of ${\cal N}=4$ SYM's. This reminds us of the KLT relation between graviton amplitudes in $\mathcal{N}=8$ supergravity and gluon amplitudes in ${\cal N}=4$ SYM \cite{Bern:2010yg, Dixon:2010gz}.\footnote{The Wilson loop-like description of MHV graviton amplitudes proposed by Brandhuber et al. \cite{Brandhuber:2008tf} may be related more directly to our line of thought. } 
 So it is tempting to think that the dual may have an interpretation as graviton scattering amplitudes in $\mathcal{N}=8$ supergravity (at strong coupling):

\medskip
We assume a simple gluing which does not break any supersymmetries.\footnote{
There is a somewhat related but different perspective one might take:  
The space we are dealing with is the two copies of $AdS_5$ with a (field theory) UV cutoff. When they are glued along the UV cutoff, it gives a realization of Randal-Sundrum geometry (RS II) \cite{Randall:1999vf}, as discussed in \cite{Chamblin:1999by}. The 4-dimensional Planck brane must be introduced at the UV cutoff, should the Israel junction condition be imposed \cite{Israel:1966rt}. 
The 4d graviton on the Planck brane is marginally localized zero modes in RS II. 
When the minimal surfaces of Alday and Maldacena are added on top of this, one might identify the graviton scattering via KLT with that of RS zero modes. Meanwhile, the RS II is conjectured to be dual to CFT coupled to 4d gravity \cite{Witten, Gubser:1999vj}. The version of RS II in \cite{Chamblin:1999by} is the double-cover of that considered in \cite{Witten, Gubser:1999vj}. So the dual of our interest may instead be two copies of CFT's coupled to 4d gravity. However, the 4d gravity cannot be of ${\cal N}=8$, since there is no known consistent coupling of ${\cal N}=8$ supergravity to ${\cal N}=4$ $SU(N>8)$ SYM's. We will not pursue this scenario here. }
The presence of the cutoff D3-branes breaks the conformal symmetry. So sixteen supersymmetries are left unbroken in each SYM theory. Since we are considering a double-copy of them, we have 32 supersymmetries in total. Thus the 4d theory has ${\cal N}=8$ supersymmetries. If the theory is local, it is plausible to expect that this is $\mathcal{N}=8$ supergravity.
The gravitons constructed by \lq\lq squaring" two gluons are different in nature from the bulk gravitons in $AdS_5$, as they are made of two independent SYM fields. So they are not dual to the energy-momentum tensor. They are coming in from and out to the asymptotic infinity of the cutoff D3-branes extended over $\mathbb{R}^{1,3}$. Thus they are asymptotic states rather than local operators in the 4d theory.
This argument suggests that the double-copy of the minimal surfaces of Alday and Maldacena may yield graviton amplitudes in 
$\mathcal{N}=8$ supergravity. However, this is very hard to check due to the strong coupling.
\subsection{The correlator/amplitude relation and KLT}
\begin{figure}[h!]
\centering
\includegraphics[height=2.8in]{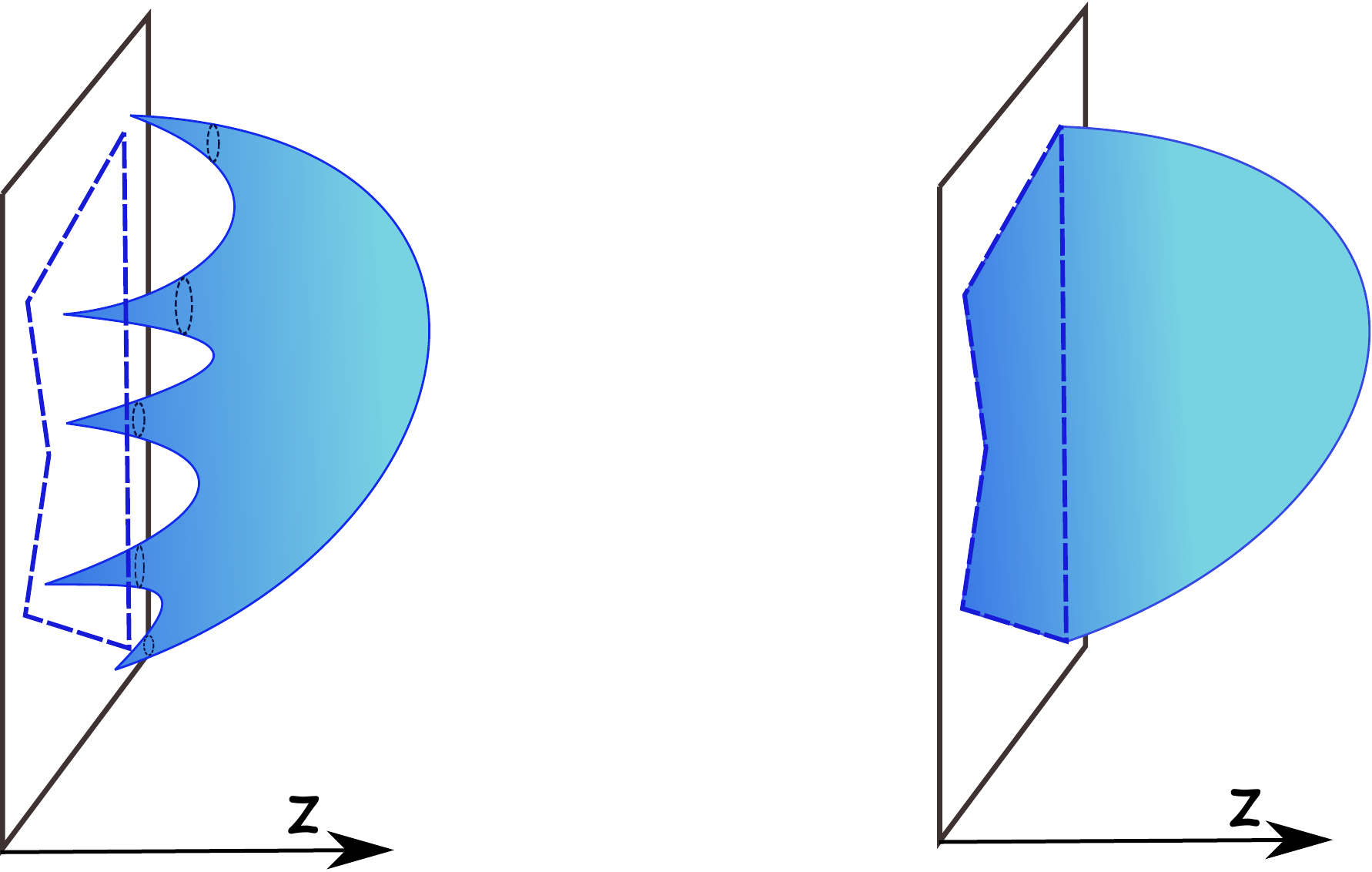}
\caption{The minimal surfaces for correlators. The right figure represents the degenerate surface after the null separation limit of the surface on the left is taken. The zigzag line forms a closed lightlike loop on the boundary.}
\label{fig.correlator}
\end{figure}

We would like to give another spin to the discussion of the KLT relation in AdS/CFT \cite{WIP}. By gluing two copies of minimal surfaces, the amplitude gets simply squared similarly to (\ref{4ptsquare}) in the flat space case. This reminds us of yet another relation, the duality between correlators and amplitudes (or Wilson loops) \cite{Alday:2010zy}. In this duality, when the separations of consecutive  adjacent operators are taken to lightlike to form a closed lightlike loop, correlation functions divided by their tree values equal the square of lightlike Wilson loops in the fundamental representation in the large $N$ limit. This relation is believed to be independent of the types of operators. This is indeed very much like the KLT relation at high energies: 

\medskip
In AdS space, correlators are described by minimal surfaces of spherical topology with spikes corresponding to operator insertions, as shown in Figure \ref{fig.correlator}. This is a stringy generalization of Witten diagram \cite{Janik:2010gc}.
It is hard to explicitly construct these surfaces, but one can expect what happens when the separations of consecutive adjacent spikes are taken to be lightlike:\footnote{We thank Tadakatsu Sakai and Masaki Shigemori for discussions on this point.} From the UV/IR relation, the curve connecting two adjacent points reaches $z\sim \Delta x$ in the bulk, where $z$ is the radial coordinate of the conformally flat Poincar\'e AdS and $\Delta x$ is the distance between the two points on the boundary. Thus in the null separation limit this curve degenerates to the straight lightlike line on the boundary. As a result, the lightlike closed loop is formed on the boundary, as in the minimal surfaces of Alday and Maldacena. This is depicted in Figure \ref{fig.correlator}. At the same time the thickness of the surface becomes zero in the limit, otherwise the area of the surface would not be minimized. Hence the closed surface in the limit is folded and becomes the double-copy of Alday and Maldacena's minimal surfaces.  We thus suggest that the correlator/amplitude duality can be thought of as an incarnation of the KLT relation.

\newpage
\vspace{1cm} \centerline{\bf Acknowledgments}

\vspace{0.5cm}
\noindent
We are grateful to Thomas S\o ndergaard, Chung-I Tan, Cristian Vergu, Giorgos Papathanasiou, Tadakatsu Sakai, and especially  Yuri Makeenko, Poul Olesen, Paolo Di Vecchia, Masaki Shigemori and Costas Zoubos for many enlightening discussions and comments on the manuscript. P.C. would like to thank the theoretical high energy physics groups at Nagoya University and Brown University for their great hospitality where a part of this work was done. 
This work was in part supported by the Grant-in-Aid for Nagoya University Global COE Program (G07).


\appendix
\setcounter{figure}{0}
\renewcommand{\thefigure}{\thesection}
\section{$N$-point kinematics}\label{Ap:Kinematics}
In order to parametrize arbitrary scattering processes with $N$ particles, it is most convenient to use generalized Mandelstam variables.  The number of independent Mandelstam variables can be counted as follows: Starting with a set of $N$ four-momenta
\begin{equation}
\left\{k_{i}^{\mu}\right\},\qquad i=1,...N\ ,
\end{equation}
we have $4\times N$ variables. Since all the particles are on-shell 
\begin{equation}
k_{i}^{2}=0\ .
\end{equation}
These give $N$ constraints. Then we have $4$ more constraints from the momentum conservation
\begin{equation}
\sum_{i}k^{\mu}_{i}=0,\qquad \mu=0,1,2,3\ .
\end{equation}
Lastly, the Lorentz rotations provide 6 more constraints. In sum, we are left with
\begin{equation}
4N-N-4-6=3N-10
\end{equation}
independent variables.

\begin{figure}
\begin{centering}
\includegraphics[scale=0.9]{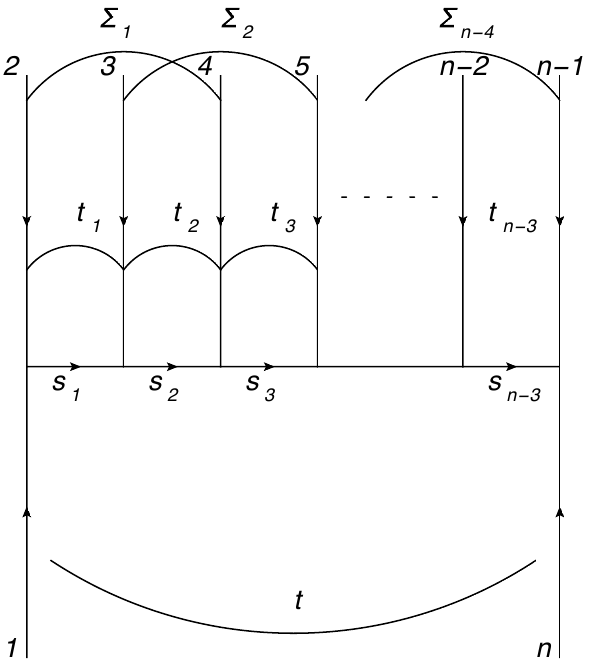}
\label{Flo:MandelstamN}
\par
\end{centering}
\caption{$N$-point Mandelstam variables.}
\end{figure}

\medskip
With this in mind, we can systematically define generalized Mandelstam variables as depicted in Figure \ref{Flo:MandelstamN}. 
For example, in the $N=4$ case, we have $3\times 4 -10=2$ independent variables. These are the center of mass energy $s$ and the momentum transfer $t$:
\begin{equation}
s=-(k_{1}+k_{2})^{2},\qquad t=-(k_{1}+k_{4})^{2}.
\end{equation}
In the case of $N=5$, there are $5\times 3 -10=5$ independent variables  given by
\begin{align}
s_{1}=-(k_{1}+k_{2})^{2},\, t_{1}=-(k_{2}+k_{3})^{2},\, t_{2}=-(k_{3}+k_{4})^{2},\, s_{2}=-(k_{4}+k_{5})^{2},\, t=-(k_{1}+k_{5})^{2}.
\end{align}
In these cases the Mandelstam variables are simply given by the sum of neighboring pairs of momenta.

\medskip
However, in the $N\ge 6$ case, it becomes a little more involved to compose the Madelstam variables.
For example, we have $6\times 3-10=8$ variables for $N=6$. There are six variables given by the sum of neighboring pairs, as one can see from  Figure \ref{Flo:MandelstamN}: 
\begin{eqnarray}
s_1=-(k_1+k_2)^2,\quad t_1=-(k_2+k_3)^2,\quad  t_2=-(k_3+k_4)^2\nonumber\\ t_3=-(k_4+k_5)^2,\quad  s_3=-(k_5+k_6)^2,\quad t=-(k_6+k_1)^2.
\label{6KIN}
\end{eqnarray}
In order to find the remaining two, it is convenient to introduce
\begin{eqnarray}
s_2=-(k_1+k_2+k_3)^2,\,\Sigma_1=-(k_2+k_3+k_4)^2,\, \Sigma_2=-(k_3+k_4+k_5)^2\ .
\label{t2S1S2}
\end{eqnarray}
In four dimensions there are at most four linearly independent momenta. Correspondingly, we have the Gramm determinant condition
\begin{equation}
\det \left(k_{i}\cdot k_{j}\right)=0\qquad (i,j=1,\ldots 5)\ .
\end{equation}
This eliminates one extra variable.

\medskip
In order to express amplitudes in terms of the generalized Mandelstam variables, it is necessary to rewrite products of non-adjacent momenta in terms of $s$, $t$ and $\Sigma$ variables.  Below we provide explicit formulas for $N=4, 5$, and $6$.

\subsection{$N=4$}
\noindent
There are only two non-adjacent products
\begin{equation}
-2k_{1}\cdot k_{3}=-2k_{2}\cdot k_{4}=-(s+t)\equiv u\ .
\end{equation}

\subsection{$N=5$ \label{KIN5}}
\noindent
There are 5 non-adjacent products, and they can be found by solving
\begin{equation}
-2k_{i}\cdot \sum^{5}_{j}k_{j}=0,\qquad i=1.\ldots 5.
\end{equation}
We find
\begin{align}
-2k_{1}\cdot k_{3}&=s_{2}-s_1-t_1,\quad -2k_1\cdot k_4=t_1-s_2-t,\quad -2k_2\cdot k_4=t-t_2-t_1\ ,\nonumber\\
-2k_2\cdot k_5&=t_2-s_1-t,\quad -2k_3\cdot k_5=s_1-t_2-s_2.
\end{align}

\subsection{$N=6$ \label{KIN6}}
\noindent
In this case we first solve \eqref{t2S1S2} to find
\begin{align}
-2k_{1}\cdot k_{3}=s_{2}-s_{1}-t_{1}\ ,\quad -2k_{2}\cdot k_{4}=\Sigma_{1}-t_1-t_2\ ,\quad -2k_{3}\cdot k_{5}=\Sigma_{2}-t_{2}-t_{3}\ .
\end{align}
Then plugging it into
\begin{equation}
-2k_{i}\cdot \sum^{6}_{j}k_{j}=0\qquad (i=1,\ldots 6)\ ,
\end{equation}
we obtain the remaining 6 non-adjacent products
\begin{align}
-2k_{1}\cdot k_4&=t_1-\Sigma_1-s_2+s_3,\quad\,-2k_{1}\cdot k_5=-t+\Sigma_1-s_3\ ,\nonumber\\
-2k_{2}\cdot k_5&=t-\Sigma_1+t_2-\Sigma_2,\quad\,-2k_2\cdot k_6= -t+\Sigma_2-s_1\ ,\nonumber\\
-2k_3\cdot k_6&=-\Sigma_2+t_3+s_1-s_2,\quad -2k_4\cdot k_6=-t_3+s_2-s_3\ .
\end{align}

\section{Saddle point equations}\label{Ap:Saddle}
In this appendix we present equations for cross-ratios that minimize the Douglas functional for $4$, $5$ and $6$ momenta. They are given by
\begin{itemize}
\item $N=4$
\begin{equation}
\frac{s}{\lambda_{1}}-\frac{t}{1-\lambda_{1}}=0\ .
\end{equation}
\item $N=5$
\begin{eqnarray}
\frac{s_{1}}{\lambda_{1}}-\frac{t_{1}}{1-\lambda_{1}}=\frac{(t-t_{1}-t_{2})\lambda_{2}}{1-\lambda_{1}\lambda_{2}}\ .\nonumber\\
\frac{s_{2}}{\lambda_{2}}-\frac{t_{2}}{1-\lambda_{2}}=\frac{(t-t_{1}-t_{2})\lambda_{1}}{1-\lambda_{1}\lambda_{2}}\ .
\end{eqnarray}
\item $N=6$
\begin{eqnarray}
\frac{s_{1}}{\lambda_{1}}-\frac{t_{1}}{1-\lambda_{1}}&=&\frac{(\Sigma_{1}-t_{1}-t_{2})\lambda_{2}}{1-\lambda_{1}\lambda_{2}}+\frac{(t+t_{2}-\Sigma_{1}-\Sigma_{2})\lambda_{2}\lambda_{3}}{1-\lambda_{1}\lambda_{2}\lambda_{3}}\ .\nonumber\\
\frac{s_{2}}{\lambda_{2}}-\frac{t_{2}}{1-\lambda_{2}}&=&\frac{(\Sigma_{1}-t_{1}-t_{2})\lambda_{1}}{1-\lambda_{1}\lambda_{2}}+\frac{(\Sigma_{2}-t_{2}-t_{3})\lambda_{3}}{1-\lambda_{2}\lambda_{3}}+\frac{(t+t_{2}-\Sigma_{1}-\Sigma_{2})\lambda_{1}\lambda_{3}}{1-\lambda_{1}\lambda_{2}\lambda_{3}}\ .\nonumber\\
\frac{s_{3}}{\lambda_{3}}-\frac{t_{3}}{1-\lambda_{3}}&=&\frac{(\Sigma_{2}-t_{2}-t_{3})\lambda_{2}}{1-\lambda_{2}\lambda_{3}}+\frac{(t+t_{2}-\Sigma_{1}-\Sigma_{2})\lambda_{1}\lambda_{2}}{1-\lambda_{1}\lambda_{2}\lambda_{3}}\ .
\end{eqnarray}
\end{itemize}



\end{document}